\documentclass{ws-mpla}
\usepackage{graphicx}
\begin{document}

\markboth{Yasuo Umino}
{Critical behavior of strongly coupled lattice
QCD at finite temperature}

\catchline{}{}{}{}{}
%
%%%%%%%%%%%%%%%%%%%%%%%%%%%%%%%%%%%%%%%%%%%%%%%%%%%%%%%%%%%%%%%%%%%%

\title{CRITICAL BEHAVIOR OF STRONGLY COUPLED\\
LATTICE QCD AT FINITE TEMPERATURE}

\author{\footnotesize YASUO UMINO}

\address{
Spain Study Center, International Programs,\\
Florida State University, Val\`encia E-46022, Spain}

\maketitle

\pub{Received (Day Month Year)}{Revised (Day Month Year)}

\begin{abstract}
We study the critical behavior of lattice Quantum Chromodynamics (QCD) in the strong coupling approximation with Kogut-Susskind and Wilson fermions at finite
temperature ($T$) and zero chemical potential. Using the Hamiltonian formulation
we construct a mean field solution to the equation of motion at finite $T$ and use
it to study the elementary thermal excitations and to extract some critical exponents characterizing the observed second order phase transition. We find similar critical behaviors for Kogut-Susskind and Wilson fermions at finite $T$.

\end{abstract}

\ccode{PACS Nos.: 03.70.+k, 12.40.$-$y}
\vspace{0.75cm}

In this letter we study the critical behavior of lattice Quantum
Chromodynamics (QCD) in the strong coupling approximation\cite{wil74} using Kogut-Susskind and Wilson fermions at finite
temperature ($T$) and zero chemical potential. This work is a natural continuation of Ref.~\refcite{umi02} where the strongly coupled lattice QCD was examined at zero $T$ but at finite chemical potential. We shall follow the same method as in Ref.~\refcite{umi02} and use the Hamiltonian formulation\cite{kog75} to
construct a mean field solution to the equation of motion at finite $T$. This solution is then used to study the elementary thermal excitations and to evaluate some critical exponents characterizing the observed second order phase transition. Comparisons of critical exponents as
well as of thermodynamic quantities indicate that Kogut-Susskind and Wilson
fermions behave similarly in the neighborhood of their critical points at finite $T$.

Analytical studies of strongly coupled lattice QCD at non zero temperature (and chemical potential) have been presented previously by several authors both in Lagrangian and Hamiltonian formulations using mostly Kogut--Susskind fermions.\cite{analytical} Where the behavior of the theory has been investigated for finite $T$ and zero chemical potential, a second order phase transition has been observed with the chiral condensate playing the role of the order parameter. The order parameter is finite for temperatures lower than some critical temperature ($T_C$) and vanishes for $T>T_C$. This is because the Lagrangians and Hamiltonians that were studied have the structure of an anti--ferromagnet with a nearest neighbor interaction, and for zero current quark mass this system is known to exhibit a second order phase transition at finite $T$.\cite{analytical} In this letter we examine more closely the critical behavior of the observed second order phase transition using both Kogut-Susskind and Wilson fermions.

As in Ref.~\refcite{umi02} our starting point is Smit's effective lattice Hamiltonian for strongly coupled QCD consisting of only quark fields.\cite{smi80} This Hamiltonian has been derived using second order strong coupling perturbation theory and only involves terms to order $\mathcal{O}(1/g^2)$ \emph{and} $\mathcal{O}(1/N_c)$, where $g$ and $N_c$ are bare quark--gluon coupling constant and the number of colors, respectively. These terms are bilinear in quark fields. Terms which are trilinear in quark fields appear at order $\mathcal{O}(1/g^4)$ for $N_c = 3$ although they are not included in the Hamiltonian of Ref.~\refcite{smi80}. We stress that our method presented here is independent of the structure of the Hamiltonian, and is therefore applicable even in the presence of these $\mathcal{O}(1/g^4)$ terms. The effective Hamiltonian is expressed in the temporal gauge supplemented by the Gauss's law constraint. This constraint restricts us to work in the subspace of the lattice (single site) Hilbert space where all physical states in the space are SU(3) color singlets and therefore locally gauge invariant.

Adopting the notation of Smit,\cite{smi80} setting the lattice spacing to unity and working in momentum space, the charge conjugation symmetric form of Smit's effective strongly coupled ($g \gg 1$) QCD Hamiltonian is
\begin{eqnarray}
\label{eqn:Hamiltonian}
H_{\rm eff}
& = &
\frac{1}{2}M_0 \left( \gamma_0\right)_{\rho\nu} \sum_{\bf p}
\biggl[ \bigl(\Psi^\dagger_{a\alpha}\bigr)_\rho({\bf p}\,),
\bigl(\Psi_{a\alpha}\bigr)_\nu(-{\bf p}\,) \biggr]_-
\nonumber \\
&   &
\!\!\!\!\!
-\frac{K}{8N_{\rm c}} \sum_{{\bf p}_1,\ldots,{\bf p}_4 }\sum_{l}
\delta_{{\bf p}_1+\cdot\cdot\cdot+{\bf p}_4,{\bf 0}}
\bigl[ e^{i(({\bf p}_2+{\bf p}_3)\cdot \hat{n}_l)}
+ e^{i(({\bf p}_1+{\bf p}_4)\cdot \hat{n}_l)} \bigr]
\nonumber \\
&   &
\!\!\!\!\!
\otimes \biggl[
\left( \Sigma_l \right)_{\rho\nu}
\bigl(\Psi^\dagger_{a\alpha}\bigr)_\rho({\bf p}_1\,)
\bigl( \Psi_{b\alpha}\bigr)_\nu({\bf p}_2\,)
-
\left( \Sigma_l \right)^\dagger_{\rho\nu}
\bigl(\Psi_{a\alpha}\bigr)_\nu({\bf p}_1\,)
\bigl( \Psi^\dagger_{b\alpha}\bigr)_\rho({\bf p}_2\,) \biggr]
\nonumber \\
&   &
\!\!\!\!\!
\otimes \biggl[
\left( \Sigma_l \right)^\dagger_{\gamma\delta}
\bigl(\Psi^\dagger_{b\beta}\bigr)_\gamma({\bf p}_3\,)
\bigl(\Psi_{a\beta}\bigr)_\delta({\bf p}_4\,)
-
\left( \Sigma_l \right)_{\gamma\delta}
\bigl(\Psi_{b\beta}\bigr)_\delta({\bf p}_3\,)
\bigl( \Psi^\dagger_{a\beta}\bigr)_\gamma({\bf p}_4\,) \biggr],
\end{eqnarray}
where $\left(\Sigma_l \right) = -i \left( \gamma_0 \gamma_l - ir\gamma_0 \right)$ with the Wilson parameter $r$ taking on values between 0 and 1. The quark field $\Psi$ is a Kogut--Susskind fermion for $r = 0$ and a Wilson fermion for $0< r \leq 1$. Using the standard summation convention color, flavor and Dirac indices for the fermion field are denoted by $(a, b)$, $(\alpha, \beta)$ and $(\rho, \nu, \gamma, \delta)$, respectively. $M_0$ is the current quark mass and the coupling constant $K$ is related to $g$ as $K = 2N_c/(N_c^2-1)\,1/g^2$. Therefore model parameters of $H_{\rm eff}$ are the Wilson parameter $r$, the current quark mass $M_0$ and the effective coupling constant $K$.

For Kogut--Susskind fermions and for vanishing current quark mass $(r = M_0 = 0)$ the effective Hamiltonian possesses a $U(4N_f)$ (global) symmetry where $N_f$ is the number of flavors. Smit has found that this symmetry is spontaneously broken down to $U(2N_f)\otimes U(2N_f)$ accompanied by the appearance of $8N_f^2$ Goldstone bosons.\cite{smi80} A finite current quark mass also breaks the original $U(4N_f)$ symmetry down to $U(2N_f)\otimes U(2N_f)$ albeit explicitly. For Wilson fermions the latter symmetry is further explicitly broken to $U(N_f)$ thereby solving the fermion doubling problem.

The effective Hamiltonian provides a description of only the ground state of strongly coupled lattice QCD. In this state there are no excited color electric flux links. Therefore the external parameter $T$ must be small enough to not to thermally excite any color electric fluxes. This condition is satisfied for $T < 1/K$.\cite{analytical}

In Ref.~\refcite{umi02} we presented an ansatz satisfying the equation of motion corresponding to $H_{\rm eff}$ for all values of $T$ and $\mu$ and explored the consequences in the $T \rightarrow 0$ limit. In this work we shall take the other limit of $\mu \rightarrow 0$. In this limit the ansatz takes on a simple form given by
\begin{eqnarray}
\label{eqn:HAAGT}
\Psi_\nu(t, {\bf p}\:)
& = &
\Bigl[ \alpha_p B({\bf p}\:) - \beta_p \tilde{B}^{\dagger}(-{\bf p}\:)\Bigr]
\xi_\nu({\bf p}\,)e^{-i\omega_p t} \nonumber \\
&    &
\;\;\;\;\;\;\;\;\;\;\; +
\left[ \alpha_p D^{\dagger}(-{\bf p}\:) - \beta_p \tilde{D}({\bf p}\:) \right]
\eta_\nu(-{\bf p}\,)e^{+i\omega_p t}.
\end{eqnarray}
In Eq.~(\ref{eqn:HAAGT}) the thermal field operators $B$ and $\tilde{B}^\dagger$ annihilates a particle and creates a hole, respectively, while $D$ and $\tilde{D}^\dagger$ are the annihilation and creation operators for an anti--particle and an anti--hole, respectively. These operators satisfy the free field fermion anti--commutation relations and each annihilation operator annihilates the \emph{interacting} thermal vacuum state $|{\cal G}(T)\rangle$. The appearance of holes and anti--holes reflects the doubling of the Hilbert space inherent in the formalism of thermo field dynamics which we have used in Ref.~\refcite{umi02}.

The coefficients $\alpha_p$ and $\beta_p$ are $\alpha_p = (1-n_p)^{1/2}$ and $\beta_p = (n_p)^{1/2}$ where $n_p$ is the Fermi distribution function $n_p= [e^{\omega_p/(k_B T)}+1]^{-1}$ with $\omega_p = [\sum_l {\rm sin}^2 ({\bf p}\cdot\hat{n}_l) + M({\bf p}\,)^2]^{1/2}$. As will be explained shortly
$M({\bf p}\,)$ is to be interpreted as a dynamical mass. It is in general both momentum and temperature dependent and is the only unknown quantity, and \emph{not} a parameter, in our ansatz of Eq.~(\ref{eqn:HAAGT}). Finally, spinors $\xi$ and $\eta$ satisfy the equation of motion of a free massive lattice Dirac Hamiltonian $H^0$ given by
\begin{eqnarray}
\label{eqn:H0}
H^0
& = &
\frac{1}{2} \sum_{{\bf p}}
\biggl[ - \sum_l {\rm sin}({\bf p}\cdot\hat{n}_l)(\gamma_0\gamma_l)_{\rho\nu}
+ M({\bf p}\,)(\gamma_0)_{\rho\nu}\biggr]
\nonumber\\
&   &
\;\;\;\;\;\;\;\;\;\;\;\;\;\;\;\;\;\;\;\;\;\;\;\;\;\;\;\;\;\;\;\;\;\;\;\;\;\;\;\;
\;\;\;\;\;\;\;\;\;\;
\otimes
\biggl[ \Psi_\rho^\dagger(t, {\bf p}\,), \Psi_\nu(t, {\bf p}\,) \biggr]_-.
\end{eqnarray}

The main idea behind our approach is to use the equation of motion for our ansatz which may be expressed as
\begin{equation}
\label{eqn:Crux}
\;\;\;\;
:\Bigl[ \bigl(\Psi_{a\alpha}\bigr)_\nu(t, {\bf p}\,), H^0 \Bigr]_- :
\:\:=\:\:
:\Bigl[ \bigl(\Psi_{a\alpha}\bigr)_\nu(t, {\bf p}\,), H_{\rm eff} \Bigr]_- :
\;\;\;\;
\end{equation}
to derive a self--consistency equation satisfied by the unknown quantity
$M({\bf p}\,)$. Here the symbol $:\hspace{0.25cm}:$ denotes normal ordering with respect to the thermal vacuum $|{\cal G}(T)\rangle$. Our solution for the strong coupling theory defined by Smit's Hamiltonian is obtained by solving the self--consistentcy equations for $M({\bf p}\,)$ for each value of $T$. With our ansatz of Eq.~(\ref{eqn:HAAGT}) obeying the free massive lattice Dirac equation, Eq.~(\ref{eqn:Crux}) describes the quark field in $H_\mathrm{eff}$ as a traditional quasi--particle with an effective, or dynamical, mass $M({\bf p}\,)$. Hence the operators $B^\dagger$, $\tilde{B}^\dagger$, $D^\dagger$ and $\tilde{D}^\dagger$ are to be interpreted as creation operators for quasi--particles, quasi--holes, quasi--anti--particles and quasi--anti--holes, respectively.

After normal ordering with the aid of computer algebra, we obtain the following
equation of motion from Eq.~(\ref{eqn:Crux})
\begin{eqnarray}
\label{eqn:EOM}
\lefteqn{ \!\!\!\!\!\!\!\!\!\!\!\!\!\!\!
\Bigl[ \sum_l \sin({\bf p}\cdot\hat{n}_l)(\gamma_0\gamma_l)
+ M({\bf p}\,)(\gamma_0) \Bigr]_{\nu\delta}
\bigl(\Psi_{a\alpha}\bigr)_\delta(t, {\bf p}\,) =
} \nonumber \\
& &
\;\;\;\;\;\;\;\;\;\;\;
\Bigl[ \sum_l A_l({\bf p}\,)(\gamma_0\gamma_l) +
B({\bf p}\,)(\gamma_0)\Bigr]_{\nu\delta}
\bigl(\Psi_{a\alpha}\bigr)_\delta(t, {\bf p}\,).
\end{eqnarray}
Thus the dynamical mass $M({\bf p}\,)$ is given by the coefficient $B({\bf p}\,)$ which is found to be
\begin{eqnarray}
\label{eqn:GAPEQ}
B({\bf p})
& = &
M({\bf p})
\nonumber \\
& = &
M_0 - \frac{3}{2}K(1 - r^2) \langle\bar{\Psi}\Psi\rangle
\nonumber \\
&   &
-\frac{1}{2}\frac{K}{N_{\rm c}} \sum_{{\bf q},l} \left( 1-2\beta_q^2 \right)
\frac{M({\bf q}\,)}{\omega_q}
\biggl[\left(1-15r^2\right)\cos({\bf q}\cdot\hat{n}_l) \cos({\bf p}\cdot\hat{n}_l)
\nonumber \\
&   &
\;\;\;\;\;\;\;\;\;\;\;\;\;\;
+ \left(1+r^2\right)\sin({\bf q}\cdot\hat{n}_l) \sin({\bf p}\cdot\hat{n}_l)\biggr]
\end{eqnarray}
where the chiral condensate $\langle \bar{\Psi}\Psi \rangle$ is given by
\begin{eqnarray}
\label{eqn:COND}
\frac{\langle \bar{\Psi}\Psi \rangle}{2V N_f N_c }
& \equiv &
\frac{1}{2V N_f N_c }
\bigl\langle\, {\cal G}(T)\bigl|
\bigl[(\bar{\Psi}_{\alpha a})_\nu, (\Psi_{\alpha a})_\nu\bigr]_-
\bigr|\, {\cal G}(T)\bigr\rangle \nonumber \\
& = &
- \sum_{{\bf p}}
\left( 1-2\beta_p^2 \right)\frac{M({\bf p}\,)}{\omega_p}.
\end{eqnarray}

We find that the ansatz of Eq.~(\ref{eqn:HAAGT}) and the equation of motion allow us to express the temperature dependent part of the vacuum energy density in the following deceptively simple form
\begin{equation}
\label{eqn:VED}
\frac{1}{2V N_f N_c }\Bigl[\langle\, H_{\rm eff}\, \rangle_{T>0} -
\langle\, H_{\rm eff}\, \rangle_{T=0} \Bigr] =
\sum_{{\bf p}}
\Biggl\{\frac{M({\bf p}\,)}{\omega({\bf p}\,)}M_0 + \omega({\bf p}\,)\Biggr\}\beta_p^2
\end{equation}
According to this definition the energy density is positive and does not depend explicitly on the strength of the interaction $K$. However we should recall that the dynamical mass that enters in Eq.~(\ref{eqn:VED}) not only depends on $K$ but also on $r$ and $T$. The expression Eq.~(\ref{eqn:VED}) forms the basis of the thermodynamics of the strongly coupled theory under investigation.

We also find the off--diagonal effective Hamiltonian to second order in thermal field operators to be
\begin{eqnarray}
\label{eqn:hoff2}
\lefteqn{ (H_{\rm eff})_{\rm off}^{(2)}|\, {\cal G}(T)\rangle = }\nonumber\\
&   &
\!\!\!\!\!\!\!\!\!\!\!\!\!\!\!
\sum_{{\bf p}} \Bigl[
\alpha_p\beta_p\, {\cal H}_1({\bf p}\,)
\tilde{B}^\dagger_{\alpha a}(-{\bf p}\,)B^\dagger_{\alpha a}({\bf p}\,)
+ \alpha_p\beta_p\, {\cal H}_2({\bf p}\,)
\tilde{D}^\dagger_{\alpha a}({\bf p}\,)D^\dagger_{\alpha a}(-{\bf p}\,)\nonumber\\
&   &
\!\!\!\!\!\!\!\!\!\!\!\!\!\!\!
+\, \alpha_p^2\, {\cal H}_3({\bf p}\,)
D^\dagger_{\alpha a}(-{\bf p}\,)B^\dagger_{\alpha a}({\bf p}\,)
+ \alpha_p\beta_p\, {\cal H}_4({\bf p}\,)
\tilde{D}^\dagger_{\alpha a}({\bf p}\,)\tilde{B}^\dagger_{\alpha a}(-{\bf p}\,)
\Bigr] |\, {\cal G}(T)\rangle
\end{eqnarray}
where ${\cal H}_i({\bf p}\,)$ are complicated functions to be presented elsewhere. In Eq.~(\ref{eqn:hoff2}) the elementary thermal excitations consist of particle--hole ($\tilde{B}^\dagger B^\dagger$), anti--particle--anti--hole ($\tilde{D}^\dagger D^\dagger$), particle--anti--particle ($D^\dagger B^\dagger$) and hole--anti--hole ($\tilde{D}^\dagger \tilde{B}^\dagger$) excitations. Note that these excitations, which carry zero total three--momenta, are color singlets and therefore satisfy the Gauss's law constraint imposed on $H_{\rm eff}$.

In the strong coupling limit the ground state of the QCD lattice Hamiltonian is any state with zero gauge flux implying that it is infinitely degenerate. Smit's effective Hamiltonian is obtained by applying perturbation theory about this ground state. The lowest nontrivial contribution comes from second order perturbation theory and  correspond to propagations of mesonic objects on the lattice. In our formulation these propagations are represented by the elementary thermal excitations in Eq.~(\ref{eqn:hoff2}) and are responsible for lifting the infinite degeneracy of $H_\mathrm{eff}$.

Thus far the results obtained using our ansatz Eq.~(\ref{eqn:HAAGT}) have been exact. We shall now invoke the mean field approximation where the equation for $M({\bf p}\,)$, Eq.~(\ref{eqn:GAPEQ}), is solved only to ${\cal O}(N_c^0)$ resulting in a self--consistency equation with a structure similar to that of the Nambu--Jona--Lasinio model in the Hartree--Fock approximation.\cite{kle92} As a consequence, the dynamical quark mass, and therefore the chiral condensate through Eq.~(\ref{eqn:COND}), become momentum independent. We note that in \emph{all} of the previous analytical studies of strongly coupled lattice QCD at finite $T$,\cite{analytical} equivalent approximations have been invoked resulting in momentum independent chiral condensates.

We shall also work in the limit of vanishing current quark mass $M_0 = 0$. In Ref.~\refcite{smi80} Smit bosonized the effective Hamiltonian, identified the quark bilinears as fundamental meson fields and showed that for Wilson fermions the pion mass vanishes when $M_0 = 6Kr^2$. This "critical" value of the current quark mass was then defined to be the chiral limit of the theory with Wilson fermions. Consequently, however, Le Yaouanc \emph{et.al.} using the same Hamiltonian as Smit claimed that above critical value of $M_0$ results in a vacuum which is not chirally degenerate, leading to the conclusion that the massless "pions" obtained by Smit are not Goldstone bosons.\cite{ley86}

The method presented here avoids the bosonization method and explicitly
constructs a solution to the equation of motion using only fermionic degrees of freedom. Furthermore, our ansatz is non-confining since it is a solution of a free massive lattice Dirac equation. Thus it does not make sense to determine the pion mass, nor masses of any other \emph{bound states}, using this ansatz. However the ansatz is not unreasonable since it exactly diagonalizes the
effective Hamiltonian to second order in field operators when $T = 0$ and $\mu \geq 0$.\cite{umi02,umi00} In addition, our formalism can accommodate a description of bound states using an improved ansatz as discussed in Ref.~\refcite{umi02}. We shall not dwell on Smit's critical value of the current quark mass for Wilson fermions until this improvement has been implemented.

\begin{figure}[t]
\begin{center}
\includegraphics[height=7cm,width=8cm]{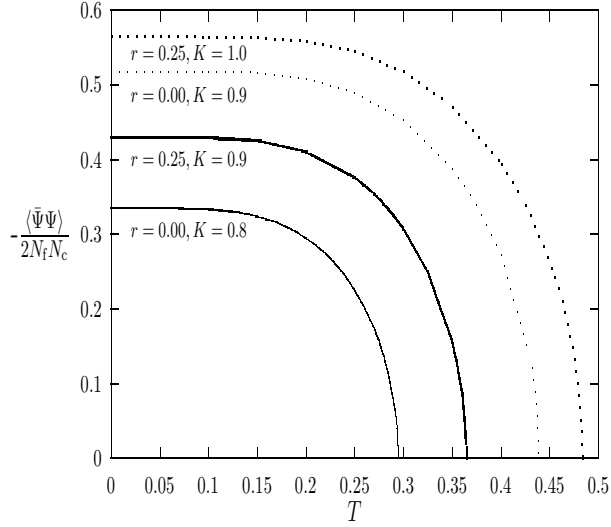}
\caption{$-\langle\bar{\Psi}\Psi\rangle/(2N_\mathrm{f}N_\mathrm{c})$ vs. $T$ for Kogut-Susskind ($r=0$) and Wilson ($r=0.25$) fermions with vanishing current quark mass $M_0=0$ and for various values of effective coupling constant $K$. Critical temperatures are shown in Table~\ref{tab:results}.}
\label{fig:ccT}
\end{center}
\end{figure}

We make the standard choice of letting the chiral condensate Eq.~(\ref{eqn:COND}) be the order parameter of the theory. In Fig.~\ref{fig:ccT} we show the behavior of $\langle\,\bar{\Psi}\Psi\,\rangle$ as a function of temperature for various sets of parameters. We observe a similar qualitative behavior between Kogut--Susskind and Wilson fermions where the chiral condensate, being finite at $T=0$, vanishes smoothly at some critical temperature $T_C$ characteristic of a second order phase transition.
For $M_0=0$, the effective Hamiltonian with Wilson fermions possesses a $U(N_f)$ symmetry for $T<T_c$ and $U(4N_f)$ symmetry for $T>T_c$. Table~\ref{tab:results} shows that for each type of fermion, $T_C$ increases with the coupling constant $K$ confirming the observations made in Ref.~\refcite{analytical}. In addition the critical temperatures shown in the table satisfy the condition $T_C < 1/K$ so that no electric fluxes are thermally excited.

\begin{table}[t]
\tbl{Critical temperature $T_C$ and coefficients $\alpha$, $\alpha '$, $\beta$ and $\eta$ for various values of input parameters. Values of $g$ corresponding to each $K$ are also shown for $N_c=3$.}
{\begin{tabular}{@{}cccccccc@{}} \toprule
  $r$ & $K$ & $g$ & $T_C$ & $\alpha$ & $\alpha '$ &$\beta$ & $\eta$ \\
  \hline
  0.00 & 0.80 & 0.97 & 0.295 & 0.023 & 0.0003 & 0.515 & $\approx$ 0.000 \\
  0.00 & 0.90 & 0.91 & 0.438 & 0.363 & 0.011 & 0.517 & $\approx$ 0.000 \\
  0.25 & 0.90 & 0.91 & 0.364 & 0.143 & 0.005 & 0.508 & $\approx$ 0.000 \\
  0.25 & 1.00 & 0.87 & 0.483 & 0.596 & 0.026 & 0.541 & $\approx$ 0.000 \\
  \botrule
\end{tabular}}
\label{tab:results}
\end{table}

The behavior of the chiral condensate near $T_C$ is best characterized by the critical coefficient $\beta$ which in this work is defined as
\begin{equation}\label{eqn:beta}
\frac{\langle\,\bar{\Psi}\Psi\,\rangle}{2N_\mathrm{f}N_\mathrm{c}} =  A_\beta(T_C-T)^\beta,\hspace{0.5cm} T_\mathrm{C} \geq T
\end{equation}
where $A_\beta$ is a constant. As shown in Table~\ref{tab:results} the critical coefficients $\beta$ are approximately equal with values close to the mean field result of $\beta = 0.5$. We may also study the response of the chiral condensate to the changes in the dynamical mass,
and define a critical exponent $\eta$ describing this response near $T \approx T_\mathrm{C}$ as
\begin{equation}\label{eqn:gamma}
\partial\frac{\langle\,\bar{\Psi}\Psi\,\rangle}{2N_\mathrm{f}N_\mathrm{c}}
/\partial M
= A_\eta(T_C-T)^{-\eta}\hspace{0.5cm} T_\mathrm{C} \geq T
\end{equation}
with $A_\eta$ being a constant. For all cases we find a linear relationship between $\langle\,\bar{\Psi}\Psi\,\rangle$ and $M$ leading to a critical coefficient of $\eta \approx 0$.

\begin{figure}[t]
\begin{center}
\includegraphics[height=7cm,width=8cm]{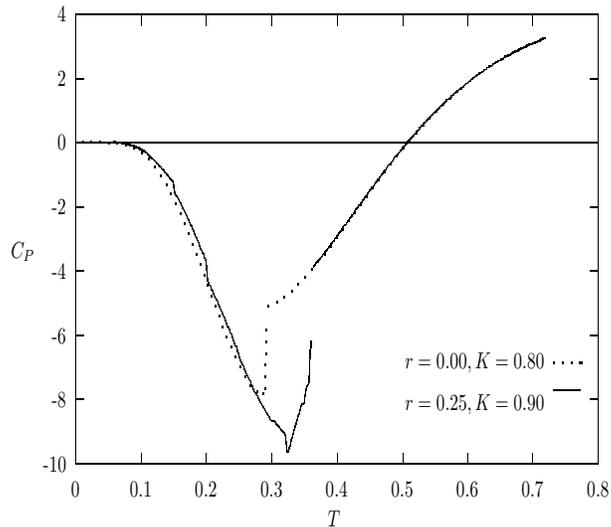}
%\vskip 8cm
\caption{Specific heat at constant pressure $C_\mathrm{p}$ vs temperature $T$ for Kogut-Susskind (r=0.0, K=0.80) and Wilson (r=0.25, K=0.9) fermions.}
\label{fig:HCT}
\end{center}
\end{figure}

As mentioned above, we may obtain all the thermodynamical information from the vacuum energy density shown in Eq.~(\ref{eqn:VED}). The qualitative behavior of the vacuum energy density for all of our model parameters are found to be the same. As an example typical behaviors for the specific heat at constant pressure $C_\mathrm{p}$ for both types of fermions are shown in Fig.~\ref{fig:HCT}. The results were obtained as follows. First the dynamical mass is determined for discrete values of $T$ by solving Eq.~(\ref{eqn:GAPEQ}) to ${\cal O}(N_c^0)$. In solving Eq.~(\ref{eqn:GAPEQ}) the mode sum is replaced by a continuum integral as has been done in Ref.~\refcite{smi80}. The resulting dynamical mass is then used to evaluate the vacuum energy density using Eq.~(\ref{eqn:VED}) followed by a polynomial fit to approximate the energy density. Finally, the interpolated energy density is used to determine the specific heat by numerically evaluating its second derivative.

We find the specific heats to be discontinuous at $T_C$. Their critical behaviors on both sides of the critical point are described by critical exponents $\alpha$ and $\alpha '$ defined as
\begin{equation}\label{eqn:alpha}
C_\mathrm{p}
= \left\{ \begin{array}{ll}
A_\alpha(T_C-T)^{-\alpha}& T_\mathrm{C} \geq T\\
A_{\alpha '}(T-T_C)^{-\alpha '} & T \geq T_\mathrm{C}
\end{array} \right.
\end{equation}
with constants $A_\alpha$ and $A_{\alpha '}$. The results for $\alpha$ and $\alpha '$ presented in Table~\ref{tab:results} vary widely, but at least they are all positive and increase with $K$ for each type of fermion.

To summarize, we began with a well defined Hamiltonian describing the ground state of strongly coupled QCD and introduced an ansatz from which we constructed a solution satisfying the equation of motion for finite $T$. Using the chiral condensate as the order parameter we find a second order phase transition for both Kogut--Susskind and Wilson fermions. In all cases we find that the critical temperature $T_C$ obeys the condition $T_C < 1/K$ so that no electric fluxes are excited near the critical point. Our ansatz allowed us to study the second order off--diagonal Hamiltonian which clearly showed the propagation of thermally excited color singlet mesonic objects responsible for lifting the infinite degeneracy of the Hamiltonian. In addition, we have extracted some critical exponents to study the critical behavior near $T_C$. Our main conclusion is that both Kogut--Susskind and Wilson fermions have similar critical behaviors at finite $T$.

Finally, as stated previously we are fully aware that our mean field ansatz Eq.~(\ref{eqn:HAAGT}) is nonconfining and needs to be improved. The need for an improved ansatz is supported by a recent numerical work by Chandrasekharan and Strouthos\cite{cha05} of strongly coupled lattice QCD at finite $T$ in which mean field theory predictions were found to fail in the immediate vicinity of the critical temperature. How to construct this improved ansatz, which goes beyond the mean field approximation and allows for the introduction of bound states, has been described in Ref.~\refcite{umi02} and shall not be repeated here. However, the method presented in this letter and in Ref.~\refcite{umi02} to introduce temperature and chemical potential is applicable regardless of the choice of the Hamiltonian and the ansatz. In particular it offers an economical way to extend the current study to the entire $T-\mu$ plane where a tricritical point is expected to be encountered.\cite{analytical}

\section*{Acknowledgments}
I thank M.--P.~Lombardo for introducing me to lattice field theory and the referee for bringing Ref.~\refcite{cha05} to my attention.

\section*{References}
\vspace*{6pt}


\begin{thebibliography}{0}
%
\bibitem{wil74}
K.G.~Wilson, {\it Phys. Rev. D} {\bf 10}, 2445 (1974).
%
\bibitem{umi02}
Y.~Umino, {\it Mod. Phys. Letts. A} {\bf 17}, 2513 (2002);
{\it Phys. Rev. D} {\bf 66}, 074501 (2002).
%
\bibitem{kog75}
J.~Kogut and L.~Susskind, {\it Phys. Rev. D} {\bf 11}, 395 (1975).
%
\bibitem{analytical}
A.~Patel, {\it Phys. Letts.} {\bf B141}, 244 (1984);
C.P.~Van Den Doel, {\it Phys. Letts.} {\bf B143}, 210 (1984);
E.--M.~Ilgenfritz and J.~Kripfganz, {\it Z. Phys.} {\bf C29}, 79 (1985);
A.~Le Yaouanc, L.~Oliver, O.~P\`{e}ne, J.--C.~Raynal, M.~Jarfi and O.~Lazrak,
{\it Phys. Rev. D} {\bf 37}, 3691 (1988); 3702 (1988);
N.~Bili\'c, K.~Demeterfi and B.~Petersson,
{\it Nucl. Phys.} {\bf B377}, 651 (1992);
X.--Q. Luo, {\it Phys. Rev. D} {\bf 70}, 091504 (Rapid Communication) (2004).
%
\bibitem{smi80}
J.~Smit, {\it Nucl. Phys.} {\bf B175}, 307 (1980).
%
\bibitem{kle92}
S.P.~Klevansky, {\it Rev. Mod. Phys.} {\bf 64}, 649 (1992).
%
\bibitem{ley86}
A.~Le Yaouanc, L.~Oliver, O.~P\`{e}ne and J.--C.~Raynal,
{\it Phys. Rev. D} {\bf 33}, 3098 (1986)
%
\bibitem{umi00}
Y.~Umino, {\it Phys. Letts.} {\bf B492}, 385 (2000).
%
\bibitem{cha05}
S.~Chandrasekharan and C.G.~Strouthos, {\it Phys. Rev. Letts.} {\bf 94}, 061601 (2005).
\end{thebibliography}
\end{document}